\documentclass[doublecol]{epl2}
\usepackage{amsmath}
\usepackage{bm}
\usepackage{amssymb}

\title{Steady self-diffusion in classical gases}
\shorttitle{Title} 

\author{J. Javier Brey\inst{1} \and M.J. Ruiz-Montero\inst{1}}
\shortauthor{F. Author \etal}

\institute{
  \inst{1} F\'{\i}sica Te\'{o}rica, Universidad de Sevilla, Apartado de Correos 1065,
E-41080 Sevilla, Spain, EU}

\abstract{A steady self-diffusion process in a gas of hard spheres at equilibrium is analyzed. The system exhibits a constant gradient of labeled particles. Neither the concentration of these particles nor its gradient are assumed to be small. It is shown that the Boltzmann-Enskog kinetic equation has an exact solution describing the state. The hydrodynamic transport equation for the density of labeled particles is derived, with an explicit expression for the involved self-diffusion transport coefficient.
Also an approximated  expression for the one-particle distribution function is obtained. The system does not exhibit any kind of rheological effects.  The theoretical predictions are compared with numerical simulations using the direct simulation Monte Carlo method and a quite good agreement is found.}

\begin{document}

\maketitle

\section{Introduction}
Self-diffusion is a particularly simple transport phenomenon in fluids that
has attracted much attention \cite{Do75,DyvB77,RydL77,McL89}. The situation usually considered corresponds to a
very dilute concentration of tagged particles \cite{Do75,DyvB77,RydL77}. Moreover, the limit of  small density gradient is considered. On the other hand, not too much attention has been devoted to study  the
peculiarities
of self-diffusion as compared with mutual diffusion when the concentrations
of the two components of the mixture are of the same order and the
density gradients are large. Also the attention paid to states exhibiting
a steady flux of tagged particles is rather restricted.

In this paper, self-diffusion  will be understood as an idealization of a
mutual two-component diffusion process in which the particles are all
mechanically identical, but some of them are assumed to be distinguishable
from the others, e.g., they carry a label, are colored, have some spin or are in different internal atomic state \cite{ChyC70}. The system considered will be a  gas at equilibrium.
It  is not assumed that the number of labeled particles is much smaller than
the total number of particles, i.e. the density of both labeled and
unlabeled particles can be of the same order. Moreover, the relative
gradient of the density of labeled particles can be arbitrarily large.

It will be shown that the kinetic
Boltzmann-Enskog equation has an exact solution describing the above state.
The kinetic theory analysis is local and does not require specification of the boundary conditions of the distribution function necessary for the state.
This is a relevant result since solutions of the Boltzmann-Enskog equation describing
nontrivial non-equilibrium states are scarce, and they provide a solid
starting point to develop a macroscopic theory of non-equilibrium states. 
Actually, the results are trivially extended to the Boltzmann equation for an arbitrary interaction potential.

\section{The state and the kinetic equation}
Consider a gas composed of mechanically identical particles of mass $m$
that is at equilibrium at temperature T, being $n$ the number of particles
density. Then, the one-particle distribution function $f({\bm v})$ of  the
system  has the form
\begin{equation}
\label{2.1}
f({\bm v})= n \varphi (v),
\end{equation}
where $\varphi (v)$ is the Maxwellian velocity distribution,
\begin{equation}
\label{2.2}
\varphi(v) = \left( \frac{m}{2 \pi k_{B} T} \right)^{d/2} e^{- \frac{m
v^{2}}{2k_{B}T}}.
\end{equation}
Here $d$ is the dimension ($2$ or $3$) of the system and $k_{B}$ is the
Boltzmann constant. 
The above distribution refers to all particles without regards to the possible existence of labels. 
Suppose now that some of the particles are labeled, Their one-particle distribution
function will be denoted by $f_{l}({\bm r},{\bm v},t)$. Since all the particles are mechanically equivalent, the equilibrium state of the
system as a whole will be conserved in time, independently of the
distribution of labeled particles.

 The Enskog equation provides a successful empirical theory to
study gases of hard particles beyond the limit of asymptotically small
density, in which the Boltzmann equation applies \cite{RydL77,McL89}.
Moreover, the low density limit of any solution of the Enskog equation is a
solution of the Boltzmann equation for hard spheres ($d=3$) or disks ($d=2$).

The state being analyzed here is characterized by a steady self-diffusion flow  of labeled particles with gradients only in the $z$-direction. Applied to this state, the Enskog equation has the form
\begin{equation}
\label{3.1}
v_{z} \frac{\partial f_{l}(z,{\bm v})}{\partial z} = g_{e}(n) J_{B}[z,{\bm
v}|f_{l}, n \varphi],
\end{equation}
where $g_{e}(n)$ is the equilibrium pair correlation function of two
particles of the gas at contact and $J_{B}$ is the Boltzmann collision
operator,
\begin{eqnarray}
\label{3.2}
J_{B} [z,{\bm v}_{1}|f_{l},n \varphi] & = &n \sigma^{d-1} \int d{\bm v}_{2}
\int d\widehat{\bm \sigma}\, \Theta \left( {\bm v}_{12} \cdot \widehat{\bm
\sigma} \right) {\bm v}_{12} \cdot \widehat{\bm \sigma} \nonumber \\
&& \times \left[ f_{l}(z,{\bm v}_{1}^{\prime})  \varphi (v^{\prime}_{2}) -
f_{l} (z,{\bm v}_{1})  \varphi( v_{2}) \right]. \nonumber \\
\end{eqnarray}
In this expression, $\sigma$ is the diameter of the particles, $d
\widehat{\bm \sigma}$ is the solid angle element around the unit vector
$\widehat{\bm \sigma}$ pointing from the center of particle $2$ to the
center of particle $1$ at contact, ${\bm v}_{12} \equiv {\bm v}_{1}-{\bm
v}_{2}$, and ${\bm v}_{1}^{\prime}$ and ${\bm v}_{2}^{\prime}$ are the
precollisional velocities given by
\begin{equation}
\label{3.3}
{\bm v}^{\prime}_{1} = {\bm v}_{1} - {\bm v}_{12} \cdot \widehat{\bm \sigma}
\widehat{\bm \sigma}, \quad \quad
{\bm v}^{\prime}_{2} = {\bm v}_{2} + {\bm v}_{12} \cdot \widehat{\bm \sigma}
\widehat{\bm \sigma}.
\end{equation}
It is worth to mention that in the present context there is no difference
between the original version of the Enskog equation \cite{ChyC70} and the
revised version introduced by van Beijeren and Ernst \cite{vByE73}. The
reason is that the local density at which the equilibrium pair correlation
must be evaluated is the total one, since this is the one determining the
collision frequency, and this density is uniform.  
The validity of the linear Boltzmann equation obtained by putting $g_{e}=1$ in Eq.\ (\ref{3.1}) to describe steady self-diffusion in the low density limit, has been discussed in detail in refs \cite{LyS82a,LyS82b}.
  Here, we will look for a normal solution to it having the form,
\begin{equation}
\label{3.4}
f_{l}(z,{\bm v})= n_{l}(z) \varphi (v) + \chi ({\bm v}) \varphi (v),
\end{equation}
where the density profile $n_{l}$  defined by
\begin{equation}
\label{2.4}
n_{l}({\bm r},t) \equiv \int d{\bm v}\, f_{l}({\bm r},{\bm v},t),
\end{equation}
is assumed to be linear,
\begin{equation}
\label{2.13}
n_{l}(z) = az+b,
\end{equation}
$a$ and $b$ being constants to be fixed by the 
hydrodynamic 
boundary conditions. To guarantee that $n_{l}$ is the actual number density of labeled particles,
the function $\chi$ introduced in Eq.\ (\ref{3.4}) must verify
\begin{equation}
\label{3.5}
\int d{\bm v} \chi ({\bm v}) \varphi (v) =0.
\end{equation}
Substitution of Eqs. (\ref{3.4}) and (\ref{2.13}) into Eq.\ (\ref{3.1})
gives
\begin{equation}
\label{3.6}
v_{z} a \varphi (v) = g_{e}(n) J_{B} [{\bm v}|\chi \varphi , n \varphi],
\end{equation}
where it has been taken into account that $\varphi$ is a steady solution of
the Enskog equation, $J_{B}[{\bm v}|\varphi,\varphi]=0$.  
Eq. (\ref{3.6}) verifies the solubility condition that the inhomogeneous term is orthogonal to the solutions of the homogeneous equation \cite{McL89,Ce75}, since the only collision  invariant in this case is unity. Moreover, Eq.\ (\ref{3.5}) establishes uniqueness of the solution. 
Because of the symmetry of the collision operator in Eq.\
(\ref{3.6}),  
its solution has to be proportional to the scalar $ a v_{z}$, i.e. 
\begin{equation}
\label{3.7}
\chi ({\bm v}) =a v_{z} \psi(v),
\end{equation}
where $\psi (v)$ is an isotropic function of the velocity. Then, the steady
flux of labeled particles is given by
\begin{equation}
\label{3.8}
j_{l,z} \equiv \int d{\bm v}\, v_{z} f_{l}(z, {\bm v})= -Da,
\end{equation}
with the self-diffusion coefficient identified as
\begin{equation}
\label{3.9}
D=- \int d{\bm v} v_{z}^{2} \psi (v) \varphi (v).
\end{equation}
Substitution of Eq.\ (\ref{3.7}) into Eq.\ (\ref{3.6})
yields
\begin{equation}
\label{3.10}
v_{z} \varphi (v) =g_{e}(n) J_{B} [{\bm v}|v_{z} \psi \varphi,n \varphi].
\end{equation}
Multiplication of this equation by $v_{z}$ and later integration over the
velocity gives
\begin{equation}
\label{3.11}
D= \frac{k_{B}T}{mg_{e} n \nu_{D}},
\end{equation}
where $\nu_{D}$ is a diffusive frequency given by
\begin{equation}
\label{3.12}
\nu_{D} = - \frac{\int d{\bm v}\, v_{z} J_{B}[{\bm v}|v_{z} \psi
\varphi,\varphi]}{\int d{\bm v}\, v_{z}^{2} \psi (v)\varphi(v)}.
\end{equation}
This expression agrees with the one obtained for the coefficient appearing
in the self-diffusion equation to Navier-Stokes order, derived from the
Enskog-Lorentz equation by means of the Chapman-Enskog procedure
\cite{McL89,BRCyG00}. It is worth to stress
that in the present context, that equation is exact for the steady state
under consideration. An
explicit evaluation of the frequency $\nu_{D}$ requires some kind of
approximation. In the first Sonine approximation, the function $\psi(v)$
becomes a constant, and it is obtained,
\begin{equation}
\label{3.13}
D=\frac{\Gamma (d/2) d}{ 4 \pi^{(d-1)/2} n \sigma^{d-1} g_{e}(n)} \left( \frac{k_{B}T}{m} \right)^{1/2}.
\end{equation}
Now, it is easy to compute the distribution function of labeled particles in the first Sonine approximation.  Equations (\ref{3.4}) and (\ref{3.7}) lead
to
\begin{equation}
\label{3.14}
f_{l}(z,v)= \left[ n_{l}(z)- c_{D}a v_{z} \right] \varphi(v),
\end{equation}
The coefficient $c_{D}$ can be expressed in terms of the self-diffusion
coefficient by use of Eq.\ (\ref{3.14}) into Eq.\ (\ref{3.8}), giving
\begin{equation}
\label{2.19}
f_{l}(z,{\bm v}) = \left[ n_{l}(z) -\frac{m a D v_{z}}{k_{B}T} \right]
\varphi (v).
\end{equation}
Define velocity moments $\mu_{k}$ by
\begin{equation}
\label{2.9}
\mu_{k} \equiv \int d{\bm v}\, v_{z}^{k} f_{l}(z, {\bm v}),
\end{equation}
$k=0,1,2, \ldots$ In particular, it is $
\mu_{0}(z) \equiv n_{l}(z)$ and    $\mu_{1} \equiv j_{l,z}\,$.
From Eq.\ (\ref{2.19}) it is obtained:
\begin{equation}
\label{2.14}
\mu_{k}(z) = c_{k} \mu_{0}(z),
\end{equation}
for $k>0$ and  even, and
\begin{equation}
\label{2.15}
\mu_{k} =-c_{k+1} \frac{a m D}{k_{B}T},
\end{equation}
for $k$ odd. Here,
\begin{equation}
\label{2.12}
c_{k} \equiv \left( \frac{2k_{B}T}{m} \right)^{k/2} \pi^{-1/2} \Gamma \left(
\frac{k+1}{2} \right).
\end{equation}
It must be noticed that Eq.\ (\ref{2.19}) has been obtained considering
the leading term in an expansion of the distribution function in orthogonal
functions. The fact that it becomes negative for large
values of $v_{z}$ is a consequence of this approximation.

\section{Simulation results}
To test the theoretical predictions presented above, computer simulations
have been performed. In order to generate the steady self-diffusion state,
the general approach initiated by Lees and Edwards \cite{LyE72} and Ashurst
and Hoover \cite{AyH73} has been employed. The basic idea is to alter the
boundary conditions in such a way that they drive the system into a
non-equilibrium steady state. The specific boundaries used here for the
self-diffusion problem are similar to those described by Erpenbeck and Wood
in ref. \cite{EyW77}.

The standard periodic boundary conditions are complemented by a color or
label change algorithm, as described now. Whenever a particle leaves the
system through the boundary located at $z=L$ ($z=0$) it is reinjected at $z=0$ ($z=L$), and
it is labeled with probability $p$ ($q$), independently of whether it was or was not labeled before.
Without loss of generality, the choice $q=1-p$ can be done in order to make
symmetric the roles of labeled and unlabeled particles. The
effect of the label change is to generate a difference in the number density
of labeled particles at both boundaries. In the long time limit, it is
expected that the system reaches a steady one-dimensional state with a
uniform flux of labeled particles along the $z$ direction.

The main goal of the simulations to be presented is to verify the existence
of the solution of the Boltzmann equation discussed above and
to check the accuracy of the derived expressions. Note that if this
solution exists, it is trivial that the corresponding solution of the Enskog
equation also exists. For this reason, the simulation technique employed has
been the direct simulation Monte Carlo (DSMC) method \cite{Bi94}. This is a
particle simulation method designed to mimic the dynamics of a dilute gas of
particles described by the Boltzmann equation. One advantage of the DSMC
method is that it allows to explote the symmetry of the system. In the
present case, we are interested in states showing gradients only in one
direction. Consequently, it is enough to consider the system as limited by
two infinite parallel walls located at z=0 and z=L, respectively, and to
divide it into layers perpendicular to the $z$-axis, being irrelevant the
coordinates of the particles perpendicular to the $z$-axis \cite{Bi94}.

In the simulations, a gas of hard spheres ($d=3$) of diameter $\sigma$  has
been employed. The reported results will be expressed in the units defined
by the mass of the particles $m$, the average mean free path $\lambda= (\pi
\sqrt{2} n \sigma^{2})^{-1}$, and the temperature $T$, taken such that
$k_{B}T =1$. Although the number of particles $N$ used in the DSMC method
does not affect the dynamics of the particles nor the physical density of
the system, and it has only a statistical meaning, let us mention for the
sake of completeness that in all the cases to be reported it has been $N/L=
800 \lambda^{-1}$.

The simulations started with the same number ($N/2$) of labeled and
unlabeled particles uniformly distributed in the system. The initial
velocity distribution of all particles was Gaussian with  vanishing average.
As expected, the system always reached a steady state, after some transient
period. Once the system was in the steady state, the relevant quantities
(hydrodynamic profiles, number of labeled particles flux, and velocity
distribution) were measured.  To identify the position dependence of the
properties, the system was divided into layers of width $\Delta z= \lambda$.
The results presented below have been averaged over a number of different
trajectories, typically 200. Also, they have been time averaged over a
period time of the order of 2000 collisions per particle.

A way of modifying the expected density gradient is to vary $p$ keeping constant the size of the system $L $.  Another possibility
is to keep constant the value of $p$ and to modify $L$. We have verified
that both methods lead to equivalent results. Then, the choice has been made of taking $p=1$, so that the boundary conditions can be interpreted as the
system being in contact with a reservoir of labeled particles at $z=L$ and a reservoir of unlabeled ones at $z=0$.

\begin{figure}
\includegraphics[scale=0.35,angle=0]{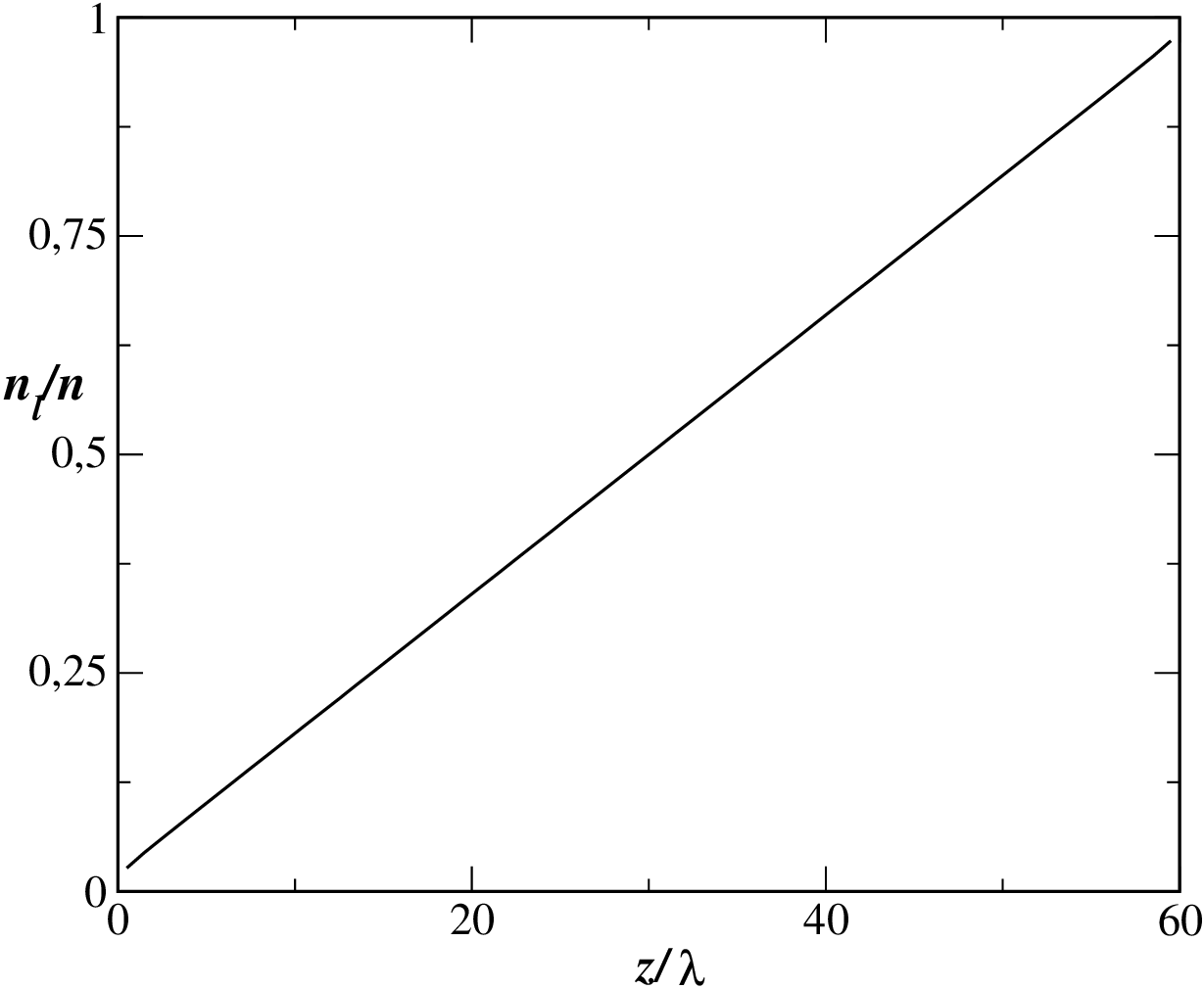}
\caption{Steady relative number density profile of labeled particles for a
system of hard spheres of width $L=60 \lambda$, where $\lambda$ is the mean
free path of the gas. \label{fig1}}
\end{figure}

\begin{figure}
\includegraphics[scale=0.35,angle=0]{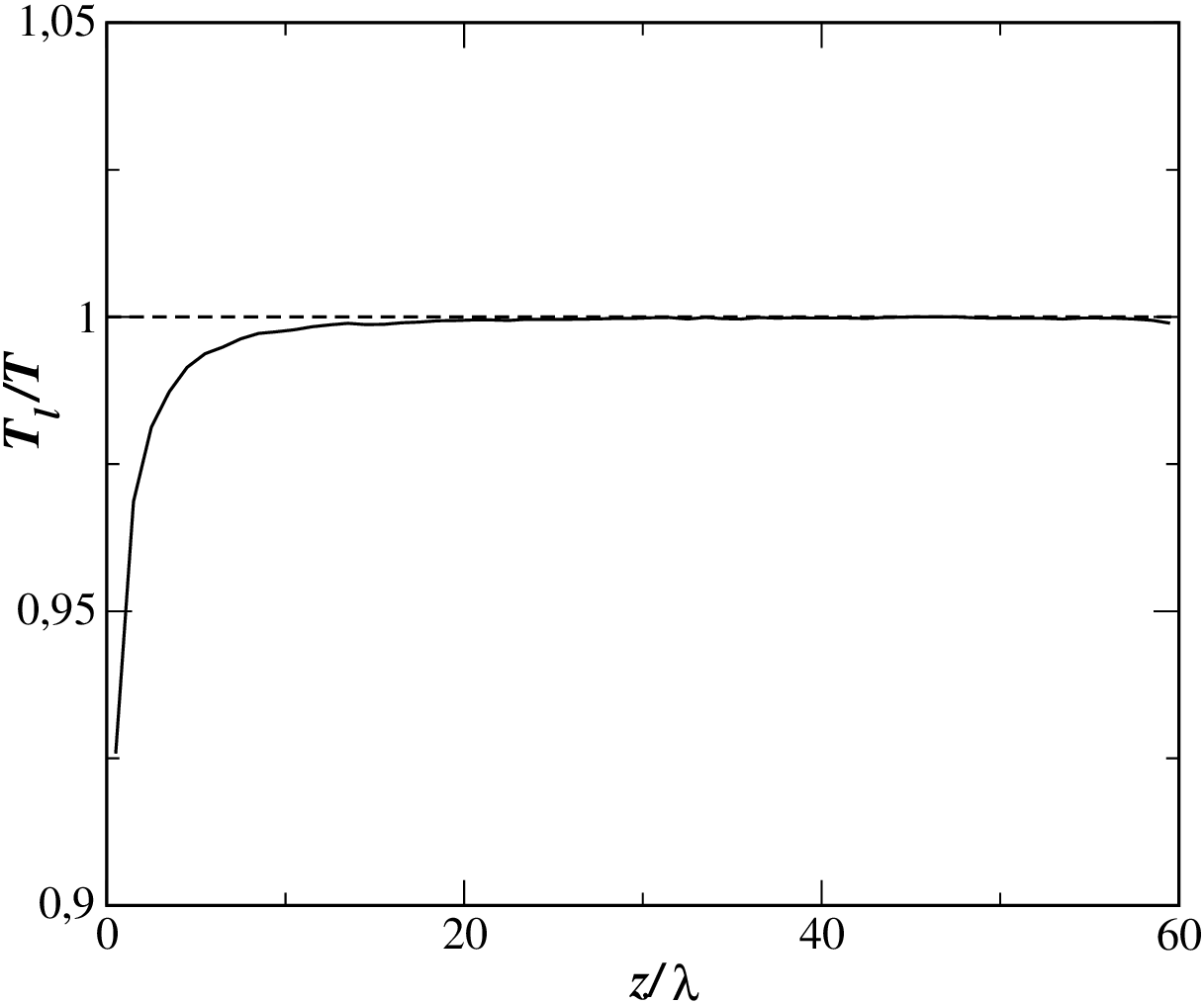}
\caption{Steady temperature profile of labeled particles $T_{l}$ for a system of
hard spheres of width $L=60 \lambda$, where $\lambda$ is the mean free path
of the gas. The temperature is scaled with the temperature of the
system $T$. \label{fig2}}
\end{figure}

\begin{figure}
\vspace{1cm}
\includegraphics[scale=0.35,angle=0]{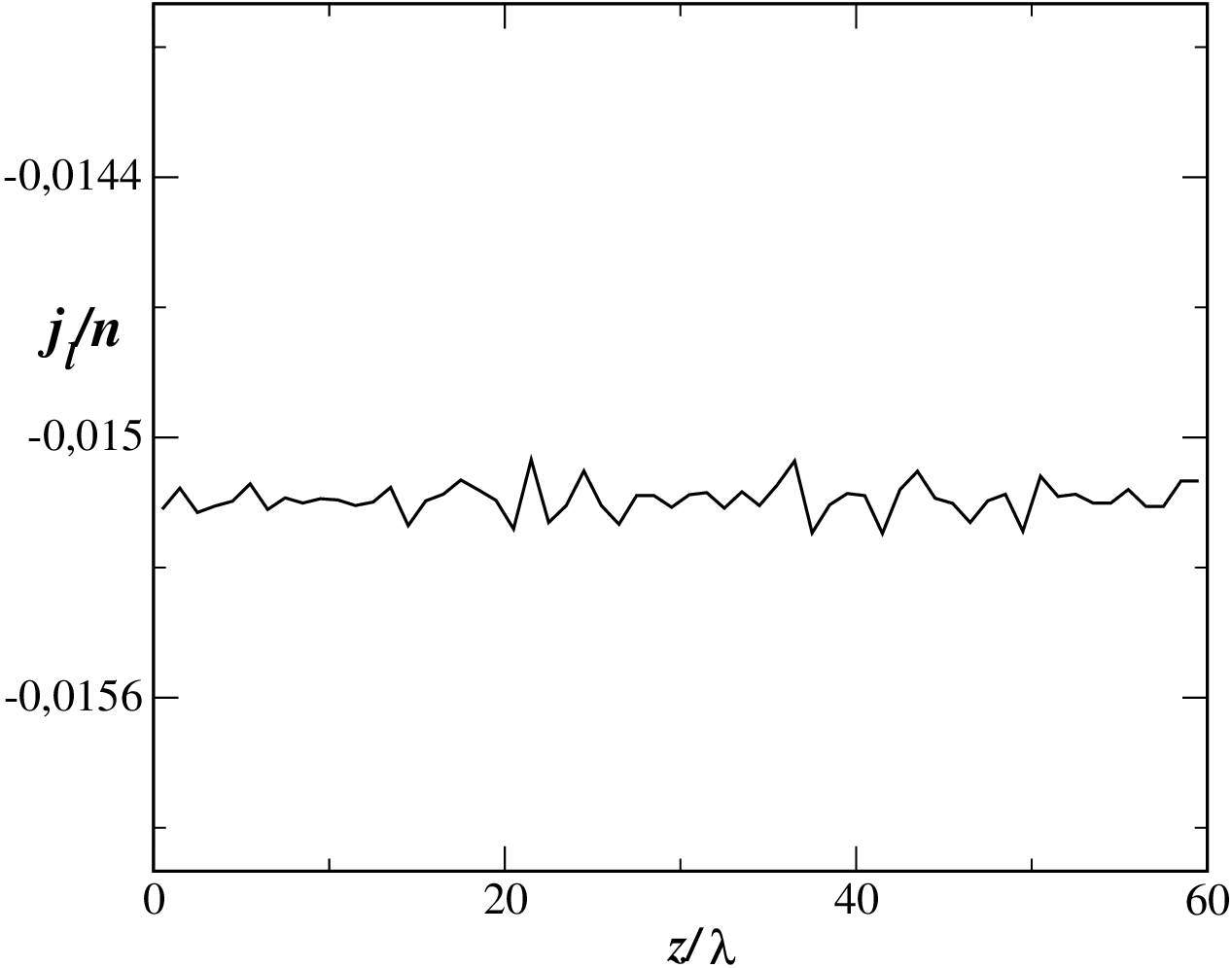}
\caption{Steady flux of labeled particles $j_{l,z}$ profile for a system of
hard spheres of width $L=60 \lambda$, where $\lambda$ is the mean free path
of the gas. The flux is measured in the dimensionless units defined in the
main text. \label{fig3}}
\end{figure}

As an example, in Figs.
\ref{fig1} and  \ref{fig2} the number density  and temperature profiles of
labeled particles in a system with $L= 60 \lambda$ are plotted. The partial
temperature of the labeled particles is defined  by considering the peculiar velocities with respect to the whole gas, which is at rest. This explains
the abrupt variation of the partial temperature near the boundary at $z=0$.
The flux of labeled particles for the same system is given in
Fig. \ref{fig3}. Its profile is uniform as it must be in a steady state, and no significant boundary layer is identified. Similar
results were  obtained in systems with $L=30 \lambda $ and $L=90 \lambda$.

The linearity of the density profiles indicates the absence of
linear rheological effects that would affect the shape of the density profile. By fitting the profiles  to a straight line, the values $a$ of the
slope have been measured. This quantity differs from $1/L$ due to the
existence of kinetic boundary layers near both walls. Then, from the
values of the current $j_{l,z}$ and $a$, the self-diffusion constant was
obtained, $ D_{sim}=- j_{l,z}/a$.

In Table \ref{t1} the computed values of the self-diffusion coefficient are
compared with the theoretical prediction $D$  given by Eq.\ (\ref{3.13}).
Since the Boltzmann dynamics is being simulated, the correlation function
$g_{e}$ has been set equal to unity. An excellent agreement is found. In
particular, the fact that the discrepancy does not show any trend with the
gradient $a$ confirms the absence of nonlinear effects predicted by the
theory. It is quite possible that the small discrepancy between theory and
simulations be due to the use of the first Sonine approximation to evaluate
the self-diffusion coefficient in the former.
\begin{center}
\begin{table}
\begin{tabular}{|c|c|c|}
\hline
$L/\lambda$ & $ a \lambda/n$ & $D_{sim}/{D}$ \\
\hline
$ 30$ & $ 3.064  \times 10^{-2}$ & $ 1.008 \pm  10^{-3}$ \\
$ 60 $ & $ 1.596 \times  10^{-2} $ & $ 1.009 \pm 2\times 10^{-3}$ \\
$ 90 $ & $ 1.0793 \times 10^{-2} $ & $ 1.008 \pm 4 \times 10^{-3} $ \\
\hline
\end{tabular}
\caption{Comparison between the self-diffusion coefficients measured in the
non-equilibrium  simulations, $D_{sim}$, and the theoretical prediction from
the Boltzmann equation, $D$, given by Eq.\ (\protect{\ref{3.13}}).
\label{t1}}
\end{table}
\end{center}

\begin{figure}
\vspace{1cm}
\includegraphics[scale=0.33,angle=0]{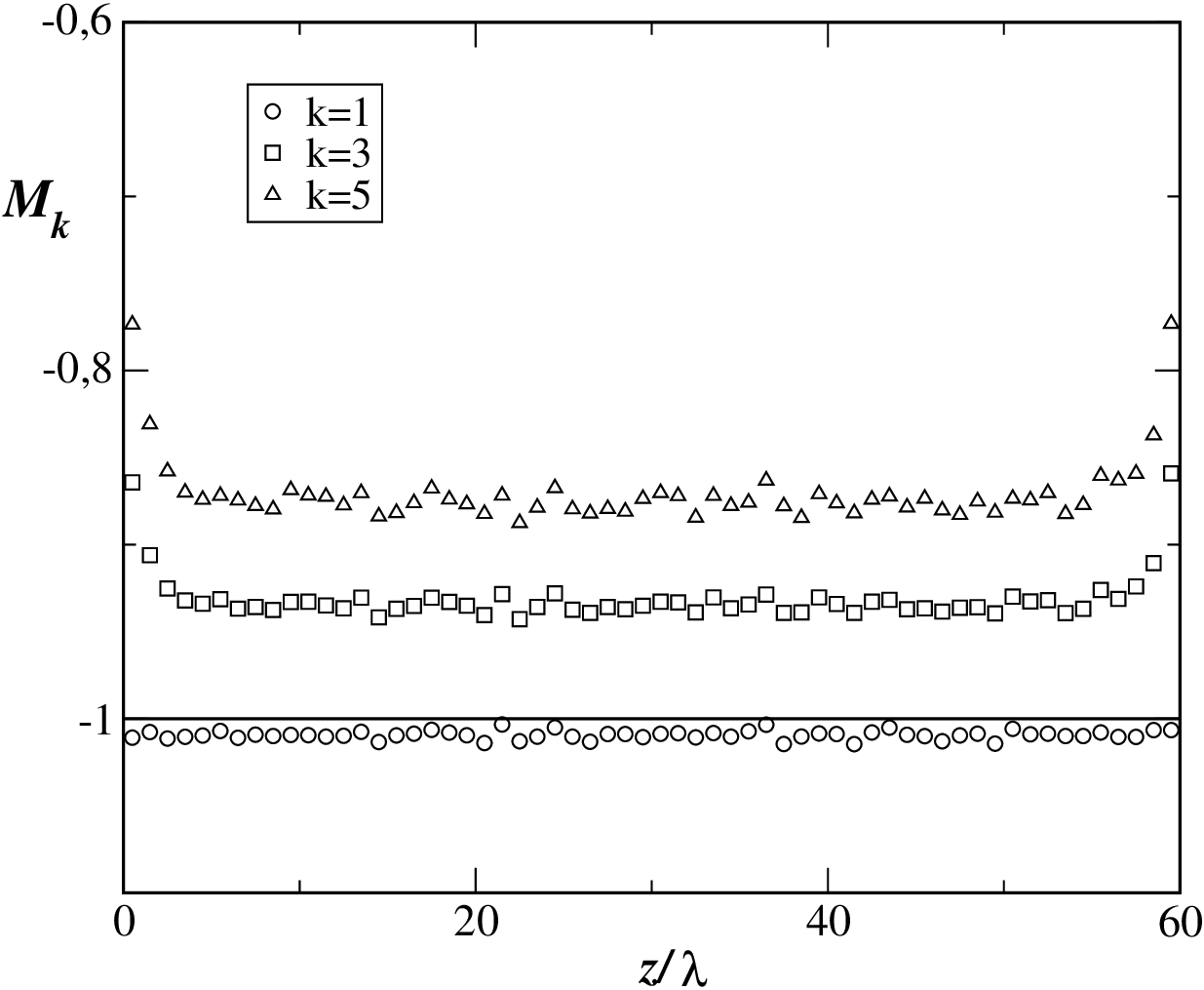}
\caption{Steady profiles of the scaled dimensionless odd moments $M_{k}$,
$k=1,3,5$, defined in Eq.\ (\protect{\ref{4.3}}), for a system with $L= 60
\lambda$. According with the theoretical prediction,  $M_{k}$ should be
uniform and equal to one for all $k$ odd. \label{fig4}}
\end{figure}

A more demanding test of the theory is to compare higher velocity moments
$\mu_{k}$, defined in Eq.\ (\ref{2.9}). Consider  first the case of $k$ being
odd. The theoretical prediction following from the Enskog  equation, and
also valid for the  Boltzmann equation,  is given by Eq.\ (\ref{2.15}). In Fig.\ \ref{fig4}, the profiles of
the scaled moments
\begin{equation}
\label{4.3}
M_{k} \equiv \frac{\mu_{k} k_{B}T}{c_{k+1} a m D}\, ,
\end{equation}
 with $k=1$, $k=3$, and $k=5$, obtained from the simulation for the system
 with $L= 60 \lambda$ are shown. It is observed that, outside rather narrow
 boundary layers, the moments are uniform along the system. Moreover, the
 values of the three moments are close, although a small systematic
 deviation from the theoretical prediction (unity)  is observed. The
 discrepancy increases as the order of the moment considered, i.e. the value
 of $k$, increases.  This seems to indicate that the origin of the
 discrepancy is the first Sonine approximation used to derive Eq.\
 (\ref{2.15}).

\begin{figure}
\vspace{1cm}
\includegraphics[scale=0.33,angle=0]{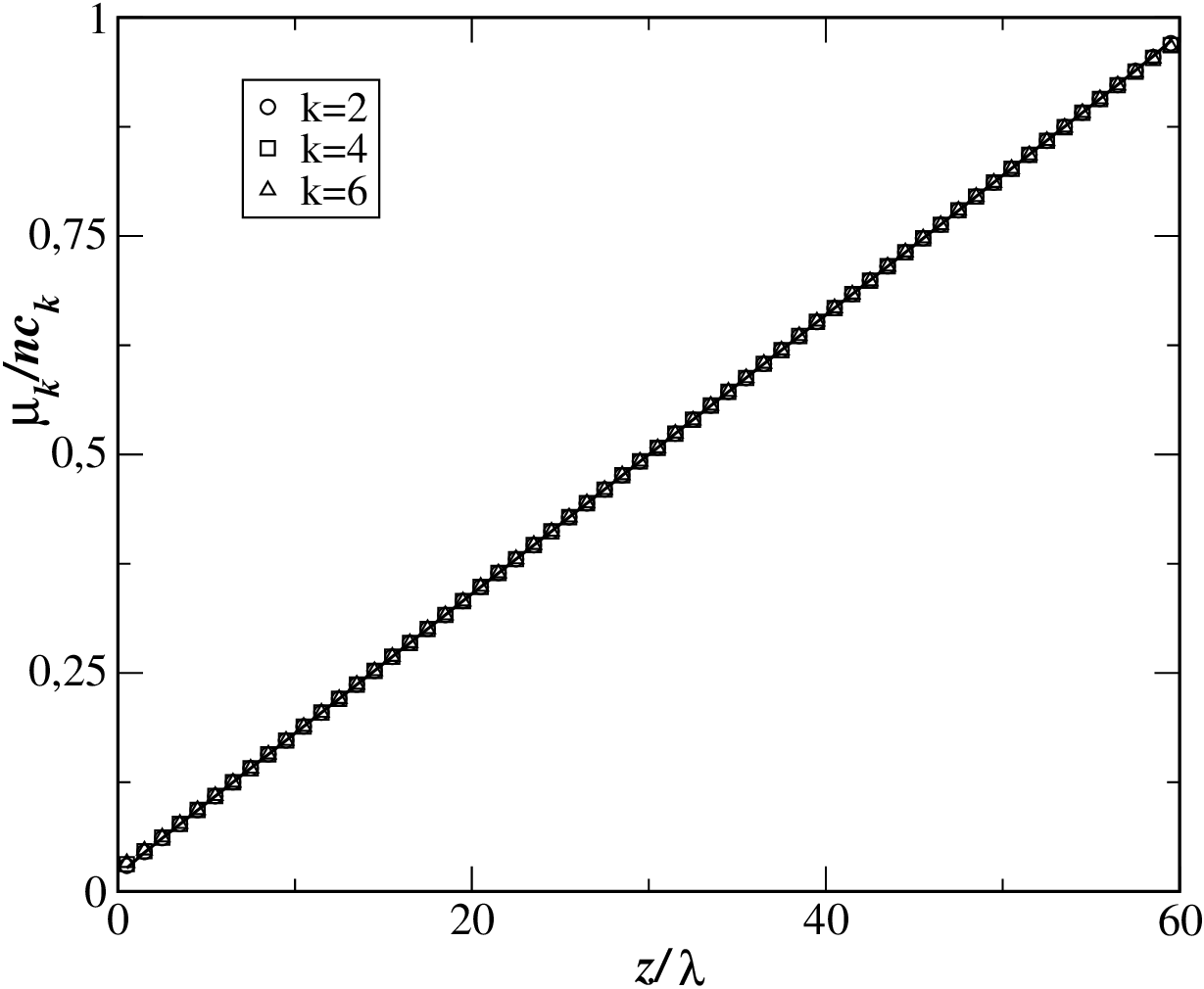}
\caption{Steady profiles of the scaled dimensionless even moments $\mu_{k}/n
c_{k}$, $k=2,4,6$, for a system with $L=60 \lambda$. Also plotted (solid
line) is the theoretical prediction, which is $n_{l}(z) /n$ for all $k$
even.   \label{fig5}}
\end{figure}

Next, consider moments $\mu_{k}$ with $k$ even. The theoretical prediction
is given by Eq. (\ref{2.14}). To test it, in Fig.\ \ref{fig5}, the measured
profiles for the ratios $\mu_{k}/n c_{k}$ for $k=2,4,6$ are plotted, again
for the system with $L= 60 \lambda$.  Now they perfectly collapse and agree
with the theoretical prediction, $n_{l}(z)/n$. The fact that the agreement
for the even moments is better than for the odd ones can be easily
understood. The even moments are determined by the first contribution to the
distribution function on the right hand side of Eq. (\ref{3.4}), and this
part is identified without resorting to the Sonine approximation.

\begin{figure}
\includegraphics[scale=0.35,angle=0]{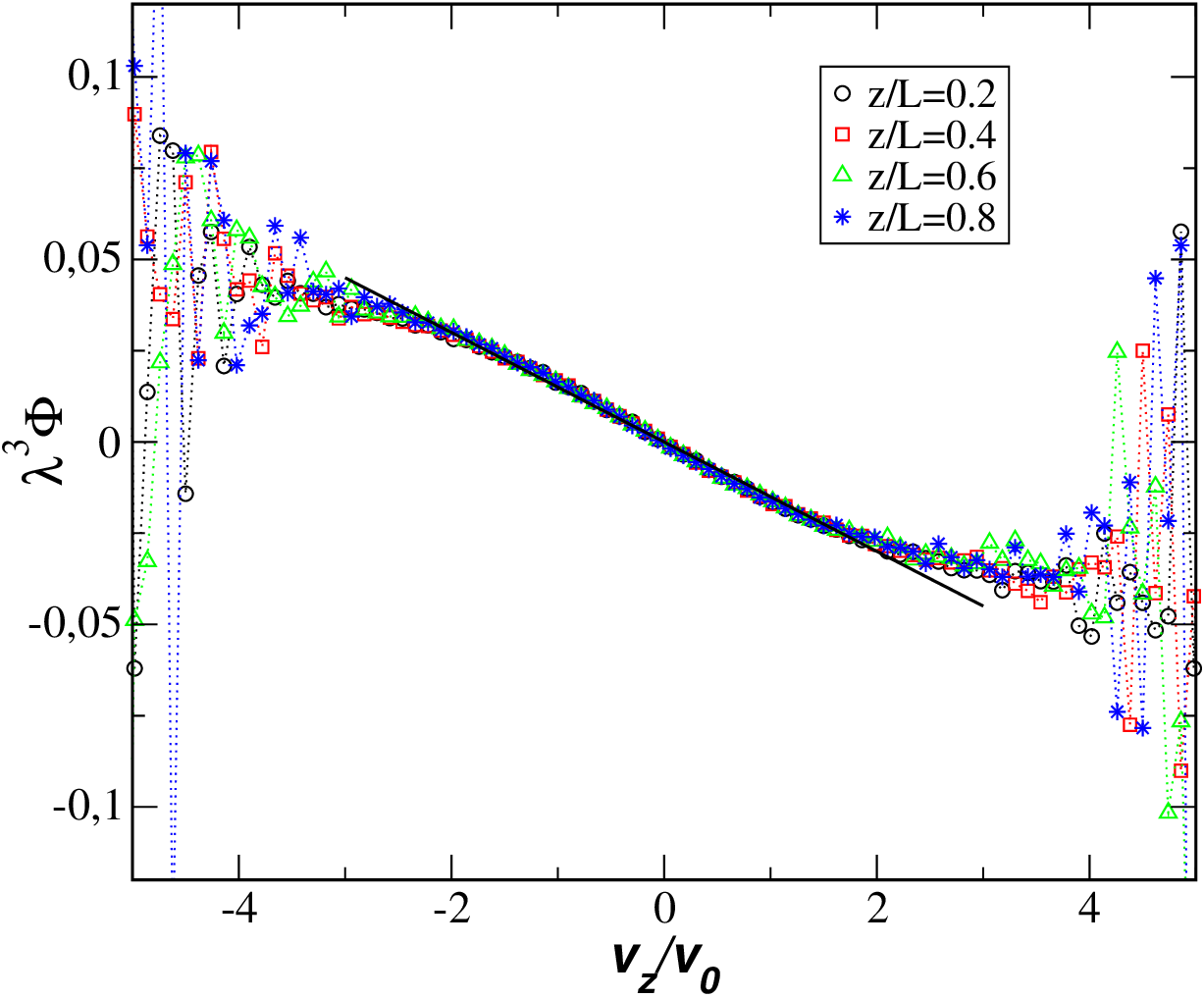}
\caption{(Color online) Marginal velocity distribution $\Phi(z,v_{z})$ for
different values of $z$ as indicated in the inset. The velocity is scaled
with the thermal velocity
$v_{0}=( k_{B}T /m)^{1/2}$. The symbols are simulation data for a
system with $L= 60 \lambda$ and the  solid line is the theoretical
prediction given by Eq. (\protect{\ref{4.6}}). \label{fig6}}
\end{figure}
Finally, let us consider the velocity distribution function itself. The
marginal distribution for the component $z$ of the velocity,
$f_{l,z}(v_{z},z)$ is defined as
\begin{equation}
\label{4.4}
f_{l,z}(z,v_{z})  = \int d{\bm v}_{\perp} f_{l}(z, {\bm v}),
\end{equation}
where the integral is carried out over those components of the velocity
perpendicular to the $z$ axis. Also define
\begin{equation}
\label{4.5}
\Phi (z,v_{z}) \equiv \frac{f_{l,z}(z,v_{z}) - n_{l}(z) \varphi_{z}
(v_{z})}{\varphi_{z} (v_{z})}.
\end{equation}
Here $\varphi_{z} (v_{z})$ is the one-dimensional Maxwell-Boltzmann velocity
distribution for $v_{z}$ with temperature $T$. The theoretical prediction
in the first Sonine
approximation follows directly from Eq.\ (\ref{3.14}),
\begin{equation}
\label{4.6}
\Phi(z,v_{z})= - \frac{mDa}{k_{B}T}\, v_{z}.
\end{equation}
i.e., it is independent of $z$ and linear in $v_{z}$. The first property is
an exact consequence of the Boltzmann equation, while the second one is
associated to the first Sonine approximation.

To measure the velocity distribution $\Phi(z, v_{z})$ in the simulations,
four slices of width $\Delta z =\lambda$, located at $z=0.2L$, $0.4L$,
$0.6L$, and $0.8L$ were considered. In Fig. \ref{fig6} the results obtained are illustrated
for the case $L=60 \lambda$. The different symbols correspond to simulation
data obtained  at different values of $z$ as indicated in the inset. The
first aspect to emphasize is that the four curves collapse, confirming that
the distribution $\Phi(v_{z})$ does not depend on $z$. Moreover, the
theoretical prediction, Eq.\ (\ref{4.6}) is also plotted, using the slope
$a$ measured in the simulations and the theoretical expression for $D$, Eq.\
(\ref{3.13}). It is seen that the agreement between theory and simulation
results is fairly good, especially in the thermal velocity region. On the
other hand, it is true that a slight curvature is clearly identified in the
simulation results. This indicates that the agreement between theory and
simulation would increase probably if higher order Sonine polynomials were
considered. Similar results were obtained for systems with $L=30 \lambda$
and $L= 90 \lambda$.

\section{Summary}
In this paper,  a steady self-diffusion process in a gas at
equilibrium has been considered. The state has been shown to be predicted  by the Enskog kinetic equation. Moreover,
DSMC simulation results obtained under appropriate boundary conditions also
indicate that the system exhibits a steady state similar to the one
described by the kinetic equation. Even more, a very good quantitative
agreement between the theoretical predictions and the simulation results has
been found. Quite interestingly, the expressions obtained for the
self-diffusion equation and for the distribution function of labeled
particles to Navier-Stkes order from the Boltzmann equation, seem to hold
also in the steady state, for arbitrary large gradients of labeled
particles.

Exact solutions of the Boltzmann and Enskog equations are rather scarce.
Moreover, self-diffusion can be considered as the prototype of transport
processes and the associate self-diffusion equation as the prototype
hydrodynamic equation. This includes not only usual molecular systems but
also
 intrinsic non-equilibrium systems, as granular gases \cite{DByL02}.
 The detailed knowledge of far from equilibrium states allows to investigate
 the way in which hydrodynamic is approached by the system, as well as many
 questions related with the stability of the state and the properties of
 hydrodynamic fluctuations far from equilibrium. In addition, the simplicity
 of the exact solution reported in this paper makes it  possible to address
 issues that are almost inaccessible otherwise.

To put the results here in a proper context, it is important to realize that
they also hold for dilute gases described by the
Boltzmann equation, with an arbitrary interaction potential. 
The Boltzmann limit of Eq. \ (\ref{3.6}) is obtained simply by putting $g_{e}(n)=1$, and the solubility condition is also verified by the Boltzmann collision operator for other interaction potentials different from hard spheres. 

As a consequence of the analysis here, some relevant questions, deserving
further consideration, arise. Given that the self-diffusion coefficient in
the steady state agrees with the Navier-Stokes one, it means that the
former
is also given by the usual Green-Kubo expression. It is then interesting
to see how it can be derived by considering the dynamics in the steady
self-diffusion state. This requires an extension of the usual linear
response theory and could be a significant first step towards
the analysis of other non-equilibrium steady states.

Another important issue is whether there exist other  self-diffusion
steady states exhibiting different density profiles of labeled particles
and, if they do, which is the relationship between them.
Also, the possible generalization to dense systems described directly by the
Liouville equation should be addressed as well as self-diffusion in steady
non-equilibrium states.

\acknowledgements

This research was supported by the Ministerio de Educaci\'{o}n y Ciencia
(Spain) through Grant No. FIS2011-24460 (partially financed by FEDER funds).


\begin{thebibliography}{0}

\bibitem{Do75}
  \Name{J.R. Dorfman}
  \Book{Fundamental Problems in Statistical Mechanics III}
  \Editor{E.G.D. Cohen}
  \Publ{North-Holland, Amasterdam}
  \Year{1975}
  \Page{277}.

\bibitem{DyvB77}
  \Name{J.R. Dorfman \and H. van Beijeren}
  \Book{Statistical Medchanics, Pt.B}
  \Editor{B.J. Berne}
  \Publ{Plenun Press, New York}
  \Year{1977}
  \Page{65}.

\bibitem{RydL77}
  \Name{P. R\'{e}sibois \and M. de Leener}
  \Book{Classical Kinetic Theory of fluids}
  \Publ{Wiley, New York}
  \Year{1977}.

\bibitem{McL89}
  \Name{J.A. McLennan}
  \Book{Introduction to Non-Equilibrium Statistical Mechanics}
  \Publ{Prentice-H, Englewood Cliffs, NJ}
  \Year{1989}.


\bibitem{ChyC70}
  \Name{S. Chapman \and T.G. Cowling}
  \Book{The Mathematical Theory of Non-uniform Gases}
  \Publ{Cambridge University Press, Cambridge}
  \Year{1970}.



\bibitem{vByE73}
  \Name{H. van Beijeren \and M.H. Ernst}
  \REVIEW{Physica A}{68}{1973}{437}.

\bibitem{LyS82a}
   \Name{J.L. Lebowitz \and H. Spohn}
   \REVIEW{J. Stat. Phys.}{28}{1982}{539}.

\bibitem{LyS82b}
    \Name{J.L. Lebowitz \and H. Spohn}
    \REVIEW{J. Stat. Phys.}{29}{1982}{39}.


\bibitem{Ce75}
  \Name{C. Cercignani}
  \Book{Theory and Application of the Boltzmann Equation}
  \Publ{Elsevier, New York}
  \Year{1975}.


\bibitem{BRCyG00}
  \Name{J.J. Brey, M.J. Ruiz-Montero, D. Cubero \and R. Garc\'{\i}a-Rojo}
  \REVIEW{Physics of Fluids}{12}{2000}{876}. In this paper, a gas of inlestic particles is considered, but the elastic limit can be easily taken.


\bibitem{LyE72}
  \Name{A.W. Lees\and S.F. Edwards}
  \REVIEW{J. Phys. C: Solid State Phys.}{5}{1972}{1921}.



\bibitem{AyH73}
  \Name{W.T. Ashurst \and W.G. Hoover}
  \REVIEW{Phys. Rev. Lett.}{31}{1973}{206}.

\bibitem{EyW77}
  \Name{J.J. Erpenbeck \and W.W. Wood}
  \Book{Statistical Mechanics, Pt. B}
  \Editor{B.J. Berne}
  \Publ{Plenum Press, New York}
  \Year{1977}.

\bibitem{Bi94}
  \Name{G. Bird}
  \Book{Molecular Gas Dynamics and the Direct Simulation of Gas Flows}
  \Publ{Clarendom Press, Oxford}
  \Year{1994}.

\bibitem{DByL02}
  \Name{J.W. Dufty, J.J. Brey, \and J.L. Lutsko}
  \REVIEW{Phys. Rev. E}{65}{2002}{051303}.



\end{thebibliography}
\end{document}